\documentclass{elsart}
\usepackage{epsfig}
\usepackage{amssymb}

\begin{document}
\begin{frontmatter}

\title{Inclusive Double Diffractive Production of SUSY Particles at the
LHC}
\author{Francis~Bursa},
\author{Agust{\'\i}n~Sabio~Vera}
\ead{sabio@hep.phy.cam.ac.uk}

\address{Cavendish Laboratory, University of Cambridge, Madingley Road, CB3 0HE, Cambridge, UK}

\begin{abstract}
We estimate the inclusive double diffractive production of SUSY particles at 
the LHC using a modified version of the POMWIG Monte Carlo event generator. 
The diffractive events are produced via the Ingelman--Schlein model for 
double pomeron exchange. The MSSM parameter space is scanned using the 
``Snowmass benchmark points'' and it is shown that the lightest Higgs boson 
is the only SUSY particle with a large 
enough rate to be detected using these diffractive events.
\end{abstract}

\end{frontmatter}

\section{Introduction}

In this Letter we investigate inclusive hard diffractive production of 
supersymmetric 
(SUSY) particles at the Large Hadron Collider (LHC) at CERN. In these 
processes 
the hard scale is provided by the mass of the centrally produced system, 
in our case the mass of the heavy SUSY particles present in the Minimal 
Supersymmetric Standard Model (MSSM). In these reactions the distinctive feature is that the protons remain 
intact after the interaction, losing only a small fraction of their initial 
energy and escaping the central detectors. The signal would be a clear one with SUSY particles tagged in the 
central region of the detector accompanied by regions of low hadronic 
activity, the so-called ``rapidity gaps''. For the study of these diffractive
interactions 
we have modified POMWIG~\cite{Cox:2000jt}, a modified version of the 
Monte Carlo event generator HERWIG~\cite{Marchesini:1991ch,Corcella:2000bw,Corcella:2002jc,Moretti:2002eu}, to 
include production of SUSY spectra. In Ref.~\cite{Cox:2001uq} POMWIG has been 
used to predict the cross--sections for double diffractive Standard Model 
Higgs and di--photon production at the Tevatron and the LHC. SUSY particle 
production has been considered in other approaches 
in~\cite{Khoze:2001xm,Kaidalov:2003fw}.

The formalism used in POMWIG to estimate diffractive cross-sections is the 
Ingelman--Schlein model for diffractive hard scattering~\cite{Ingelman:1984ns}. In this model the interaction is triggered by a ``double pomeron exchange'' 
and the production cross--section factorises into a product of a 
Regge flux factor and a parton distribution function. If the concept of 
Regge factorisation is to a good approximation 
universal then it can be applied to the present study using 
the diffractive parton distributions measured in deep inelastic scattering 
experiments at HERA, where this model has proved to be 
successful~\cite{Adloff:1997sc}. For 
the large centre--of--mass energy of the LHC the only 
Regge exchange needed to be taken into account is pomeron exchange, this is 
done using a pomeron flux factor $f_{~{\rm P \hspace{-0.25cm}I}~/p}$ and a 
pomeron parton density $g(x,m_{\rm X}^2)$. 
In POMWIG the pomeron flux is parameterised as
\begin{eqnarray}
f_{~{\rm P \hspace{-0.25cm}I}~/p} \left(x_{\,{\rm P \hspace{-0.25cm}I}~}\right)
&=& \int^{t_{\rm min}}_{t_{\rm max}}
\frac{e^{B_{\hspace{-0.06cm}~{\rm P \hspace{-0.25cm}I}~} \,t}}{x_{\,{\rm P
\hspace{-0.25cm}I}}^{2\,\alpha_{\hspace{-0.06cm}~{\rm P
\hspace{-0.25cm}I}~}\,(t)-1}}
\end{eqnarray}
with $x_{~{\rm P \hspace{-0.25cm}I~}}$ being the proton's energy fraction 
carried by the pomeron, $t$ the proton momentum transfer, $B_{\,{\rm P
\hspace{-0.25cm}I}~} = 4.6$ the diffractive 
slope and $\alpha_{\,{\rm P \hspace{-0.25cm}I}~}(t) = 1.20 + 0.26 \, t$ the 
pomeron trajectory. For details on the choice of these values see 
Ref.~\cite{Cox:2000jt}. This approach works well for the description of 
dijets at the Tevatron~\cite{Appleby:2001xk}.

A theoretical uncertainty in the present estimates stems from the fact that, 
in processes where the 
incoming beam particles have hadronic structure, the rapidity gaps can be 
filled due to secondary interactions spoiling 
the clean signal~\cite{Dokshitzer:1991he,Bjorken:1992er}. This  
affects the prediction for the cross--sections by a normalisation factor 
mainly depending on the centre--of--mass energy. Based on recent estimates for 
LHC energies it is possible to take into account these effects by multiplying 
the obtained cross--section by a gap survival probability factor 
of~$\sim 0.02 - 0.026$~\cite{Khoze:2000cy,Kaidalov:2003fw}.

In this Letter the focus is on double diffractive collisions of the form  
$p+p \rightarrow p+{\rm gap}+X+{\rm gap}+p$, where $X$ represents the decay 
products of the SUSY particles 
and some pomeron remnants. To have a diffractive 
process the energy fraction lost by the incoming hadrons, which we call $\xi$, 
should be smaller than $\xi_{\rm max} = 0.1$. Ideally, proton tagging 
detectors in the forward and backward directions would be needed to take 
full advantage of these signals and to be able to reconstruct the masses of 
the new particles. The analysis of diffractive collisions is experimentally 
challenging, even at medium luminosity at the LHC the diffractive events 
would be contaminated by other non--diffractive interactions taking place 
in the same bunch crossing. In principle, to reconstruct the gap in the hard 
subprocess, it would be possible to use tracking subdetectors, for a 
discussion on this issue see Ref.~\cite{DeRoeck:2002hk}. In this work only 
signals are estimated, leaving the calculation of possible backgrounds 
for a future publication.

The paper is organised as follows: In Section~2 we reproduce previous 
results in the literature regarding Standard Model Higgs production. In 
Section~3 we study the production of SUSY particles to conclude that only 
the lightest Higgs boson has large enough cross--sections. In Section~4 
we study its production at the different benchmark points characterizing 
the MSSM parameter space. In Section~5 we present our conclusions.

\section{Standard Model Higgs Production}
\label{SMH}

In this Section we reproduce some of the results in Ref. \cite{Cox:2001uq} 
for the double diffractive production of the Standard Model Higgs 
at the LHC. In this 
way we explain the methodology which will later be used in the SUSY case. 
We set the  mass of the Higgs boson to be 115 GeV. In double diffractive Higgs production the total cross--section is 
calculated for $\xi<\xi_{\rm max}$ and reads
\begin{eqnarray}
\sigma &\simeq& \frac{G_F \alpha_s^2}{288 \pi \sqrt{2}}\frac{m_h^2}{s} 
\int_{{m_h^2}/{s}}^1 \frac{dx}{x} \, g_1(x,m_h^2) \, 
g_2(\frac{m_h^2}{s\,x},m_h^2)
\end{eqnarray}
where $\sqrt{s}$ is the hadron--hadron centre--of--mass energy, and 
\begin{eqnarray}
g_i (x,Q^2)&=& \int_x^{\xi_{\rm max}} d\xi_i \, 
f_{~{\rm P \hspace{-0.25cm} I~}/i}(\xi_i) 
\, g_{~{\rm P \hspace{-0.25cm} I}} \left(\frac{x}{\xi_i},Q^2\right)
\end{eqnarray}
a convolution of the pomeron flux and a parton distribution in the 
pomeron. 

To evaluate the differential cross--sections ${d \sigma}/{d \xi}$ and 
${d \sigma}/{d \beta}$, with $\beta$ being the fraction of the pomeron 
momentum carried by the gluon, we work with an updated version of POMWIG 
using HERWIG version 6.5, which includes SUSY hard subprocesses, to generate 
diffractive interactions. The incoming particles are set to be protons with 
an energy of 7000 GeV. From the generated events we select those for 
which $\xi<0.1$ for both incoming protons, and extract the values of the 
variables $\beta_i$ which are the ratios of gluon momentum to pomeron 
momentum for the pomeron radiated by proton $i$. 

To calculate the differential cross--section ${d^2 \sigma}/{d\xi d\beta}$ 
the weight of each event is added to the appropriate bin in 
$\left(\xi,\beta\right)$ space. Half of each event's weight is added 
to the bin corresponding to its values of $\xi_1$ and $\beta_1$, and half 
to the bin corresponding to its values of $\xi_2$ and $\beta_2$. This has 
the effect of symmetrising the differential cross--section:
\begin{eqnarray}
\frac{d \sigma}{d \xi d \beta} &=& \frac{1}{2} \left(\int_0^{0.1} 
d \xi_1 \int_0^1 d\beta_1 \frac{d\sigma}{d \xi_1 d\xi_2 d\beta_1 d\beta_2}+
\int_0^{0.1} 
d \xi_2 \int_0^1 d\beta_2 \frac{d\sigma}{d \xi_1 d\xi_2 d\beta_1 d\beta_2}
\right).
\end{eqnarray}
At the end, this expression is summed over all $\beta$ or $\xi$ to obtain 
the single--differential cross--sections. The results are shown in Fig. 
\ref{SMH}, they are consistent with those in Ref. \cite{Cox:2001uq} 
\footnote{We take the results as given by HERWIG 6.5 and, differently to 
Ref. \cite{Cox:2001uq}, we do not double the cross--section to estimate the 
effects of NLO QCD corrections.}. 
\begin{figure}
\begin{center} 
\includegraphics[width=4.8cm, angle=-90]{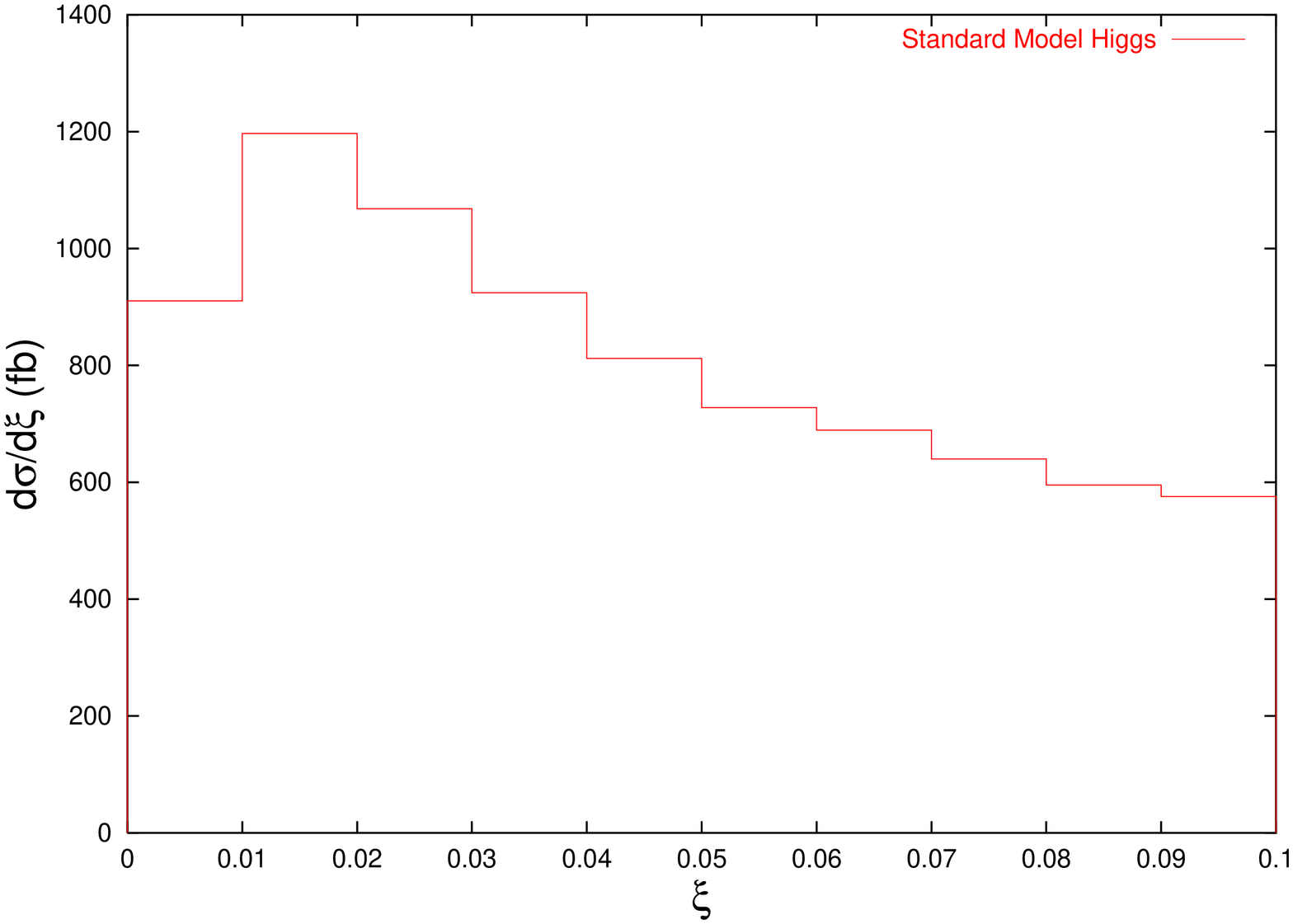}\includegraphics[width=4.8cm,
angle=-90]{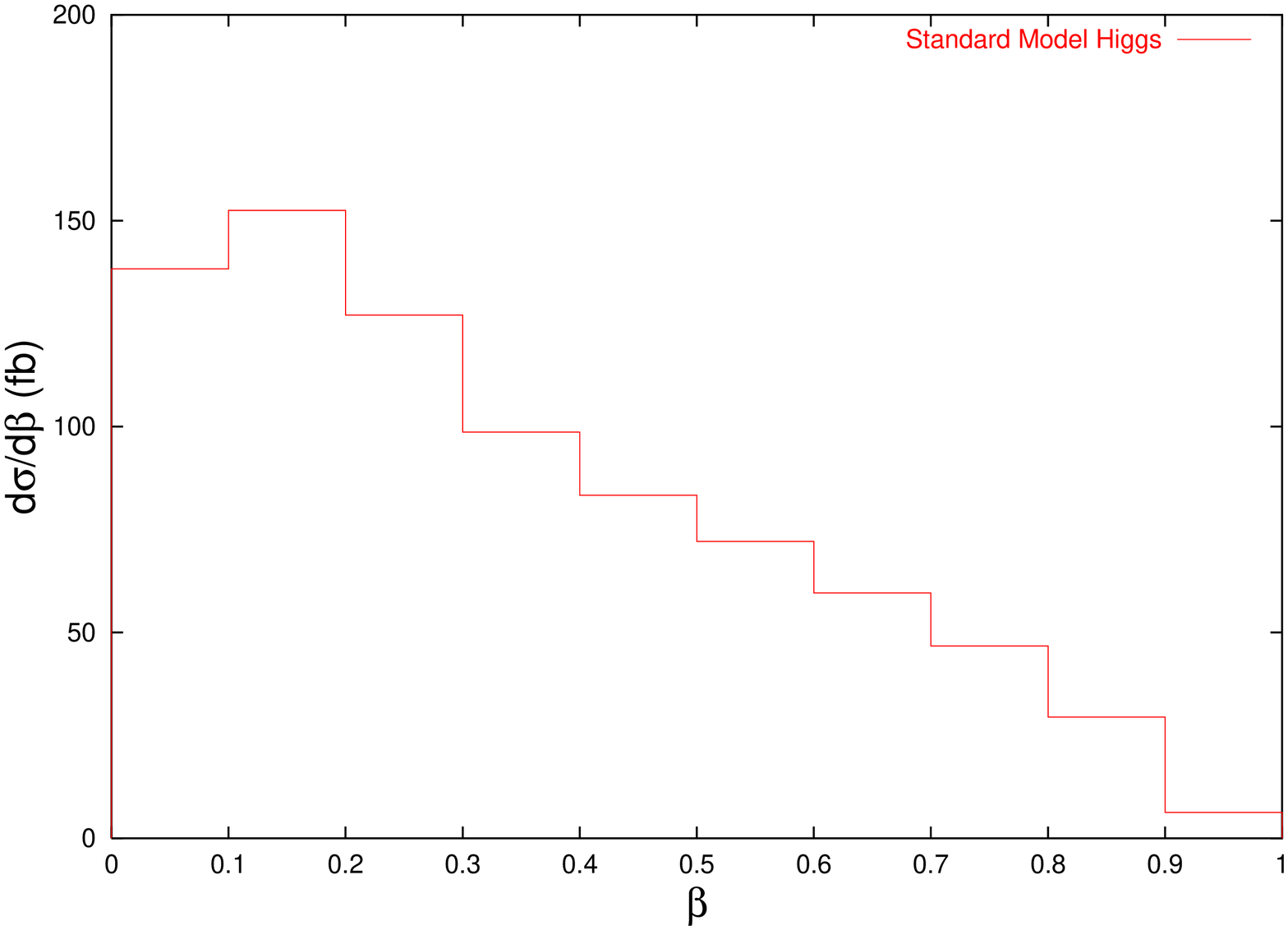} 
\caption{Differential cross-sections for inclusive double diffractive 
production of the Standard Model Higgs at the LHC.}\end{center}
\label{SMH}
\end{figure}

The inclusive double diffractive cross--section 
is $81.4 \pm 1.0$ fb, which is consistent with the result obtained in 
Ref. \cite{Cox:2001uq} when the POMWIG default fit of the H1 pomeron structure 
function was used. The fact that the distributions are not forced to 
high values of $\xi$ and $\beta$ implies that at the LHC the hadron--hadron 
centre--of--mass energy is large enough to easily generate a gluon--gluon 
centre--of--mass energy squared larger than the square of the mass of the
produced 
particle, i.e. $\hat{s} = s \xi_1 \xi_2 \beta_1 \beta_2 > m_h^2$. We will later 
see that this is also true in the SUSY case. The final result 
should be corrected to include the gap survival factor which, from theoretical 
considerations, at 14 TeV would be of order $2\%$. 

Given these results for the Standard Model Higgs in the next Section we 
estimate what the cross--sections would be in the case of the MSSM. 

\section{Supersymmetric particles production}
\label{SUSYp}

Once the Standard Model results have been obtained, to study the 
SUSY processes of interest, we should specify the regions of 
the MSSM parameter space we want to 
investigate. To define the masses, couplings and decay modes for 
the SUSY particles, we use the so--called ``Snowmass Points and Slopes'' 
(SPS), a set of benchmarks for SUSY searches. In Ref.~\cite{Allanach:2002nj}
 an unconstrained 
version of the MSSM is proposed where all possible soft SUSY breaking terms 
are added to the Lagrangian and then different parameterisations of these 
terms are considered. As is well known, the number of free parameters in 
the theory is very large but it can be reduced if a particular SUSY breaking 
(SB) mechanism is assumed. The most popular ones are minimal 
supergravity (mSUGRA), gauge--mediated SUSY breaking (GMSB), and anomaly 
mediated SUSY breaking (AMSB) (for a brief description of these scenarios 
see, for example Ref.~\cite{Allanach:2002nj}). These SB scenarios have a 
reduced three or four dimensional parameter space.

We have calculated the diffractive production of all neutral MSSM 
Higsses ($h^0,H^0,A^0$), charged Higsses, gauginos, spartons and sleptons. 
The cross--sections for production of SUSY particles other 
than the lightest SUSY Higgs, $h^0$, are small and, at least for 
these SPS benchmarks, it renders the inclusive double diffractive channel 
as not an optimal one to study them (of all the cross--sections studied the 
second largest is that of squark production where even pushing the parameter 
space to low squark masses the cross--section is 
$\sim {\mathcal O} (40~$fb)).  
Although we are investigating double diffractive production in this Letter, 
it would be interesting to study if the rates of production for these SUSY 
heavy states are higher in other processes like single diffractive 
production. The situation for $h^0$ is far more 
positive. The production cross--sections, which are dominated by the gluon 
exchange channel, are larger than for the 
Standard Model case. We will show this in the next Section where we also 
include a brief description of the different MSSM benchmark points.

\section{Inclusive Double Diffractive Production of MSSM lightest Higgs}

In this Section we show the results for the production of the lightest 
MSSM Higgs. The analysis proceeds in the same way as in Section 2. For 
completeness we write down the matrix element used by HERWIG for the 
gluon--gluon $\rightarrow$ $h^0$ hard subprocess:
\begin{eqnarray}
&&\hspace{-1cm}\overline{\left|M\right|^2} 
= \frac{\alpha_{\rm em} \alpha_s^2 m_{h^0}^4}{72 \pi \sin^2{\theta_W}
\left(N_c^2-1\right)m_W^2} \left|\sum_{\rm f} g_f^{h^0} A_f^{h^0}
\left(\frac{4 m_f^2}{m^2_{h^0}}\right)+\sum_{\rm {\tilde f}} 
g_{\tilde f}^{h^0} A_{\tilde f}^{h^0}\left(\frac{4 m_{\tilde
f}^2}{M^2_{h^0}}\right)\right|^2 
\end{eqnarray}
where the sum over fermion includes the loops of heavy quarks ($b$ and $t$) 
and the sum over sfermions takes into account loops with squarks (${\tilde b}$
and ${\tilde t}$). The expressions for the coefficients $g$ and $A$ can be 
found in Ref. \cite{Spira:1997dg}.\\
\\
{\bf The Snowmass points}\\
\\
We now give a very brief description of the Snowmass points used to scan 
the MSSM parameter space. For each of the points we have calculated the total 
and differential cross--sections for inclusive double diffractive production 
of the lightest MSSM Higgs. The value of the top--quark mass in all the 
SPS benchmark scenarios is 175 GeV and the sign of the $\mu$--term in the 
superpotential is taken to be positive. The mass of the Higgs is kept close to 
115 GeV (we show the exact values for each of the SPS points in the tables below) and the spectra for the other SUSY particles for all the 
benchmark points used here can be found in Ref. \cite{Allanach:2002nj}. 
We again remind the reader that the values for the cross--sections should 
include a correcting factor to include gap survival probability effects.\\
\\
{\bf SUSY breaking in minimal supergravity (mSUGRA SPS 1-5)}\\
\\
In these scenarios the breaking of SUSY takes place in a hidden sector and 
is mediated to the visible MSSM sector via gravitational interactions. This 
proposal is parameterised by a scalar mass $m_0$, a gaugino mass $m_{1/2}$, 
a trilinear coupling $A_0$ and 
the ratio of the vacuum expectation values of the 
two Higgs doublets $\tan{\beta}$.

The values of the parameterisation for these scenarios are given 
in Table~1. The differential cross--sections obtained using the 
SUSY version of POMWIG are 
shown in Fig.~2. These plots show that the cross--sections are larger 
than the corresponding ones in the Standard Model. As can be seen in Table~1 
the value of the diffractive total cross--sections ranges from 
92 fb to 190 fb. These results are very similar for the rest 
of the MSSM points showing that the number of events would be large at the 
energies delivered at the LHC and, if the backgrounds are not very large, 
this diffractive channel would be an interesting one to identify the MSSM 
lightest Higgs.
\begin{table}[h] 
\begin{center}
    \begin{tabular}{|c|c|c|c|c|c|c|c|c|c|}  \hline      
    SUGRA&$m_0$&$m_{1/2}$&$A_0$&$\tan{\beta}$&$m_{h^0}$&$\sigma_{~{\rm P
\hspace{-0.25cm}I}}$ (fb)\\
    \hline\hline       
    $1a$&100&250&-100&10&114&190\\ 
    $1b$&200&400&0&30&118&167\\ 
    $2$&1450&300&0&10&116&175\\ 
    $3$&90&400&0&10&117&171\\ 
    $4$&400&300&0&50&115&184\\ 
    $5$&150&300&-1000&5&120&92\\ 
    \hline     
\end{tabular}
\label{SUGRA}
\vspace{0.4cm}
\caption{Parameterisation of mSUGRA points and total diffractive 
cross--sections.}
\end{center}
\end{table}
\begin{figure}
\label{mSUGRA}
\begin{center} ~~~~~~mSUGRA\\
\includegraphics[width=4.8cm,
angle=-90]{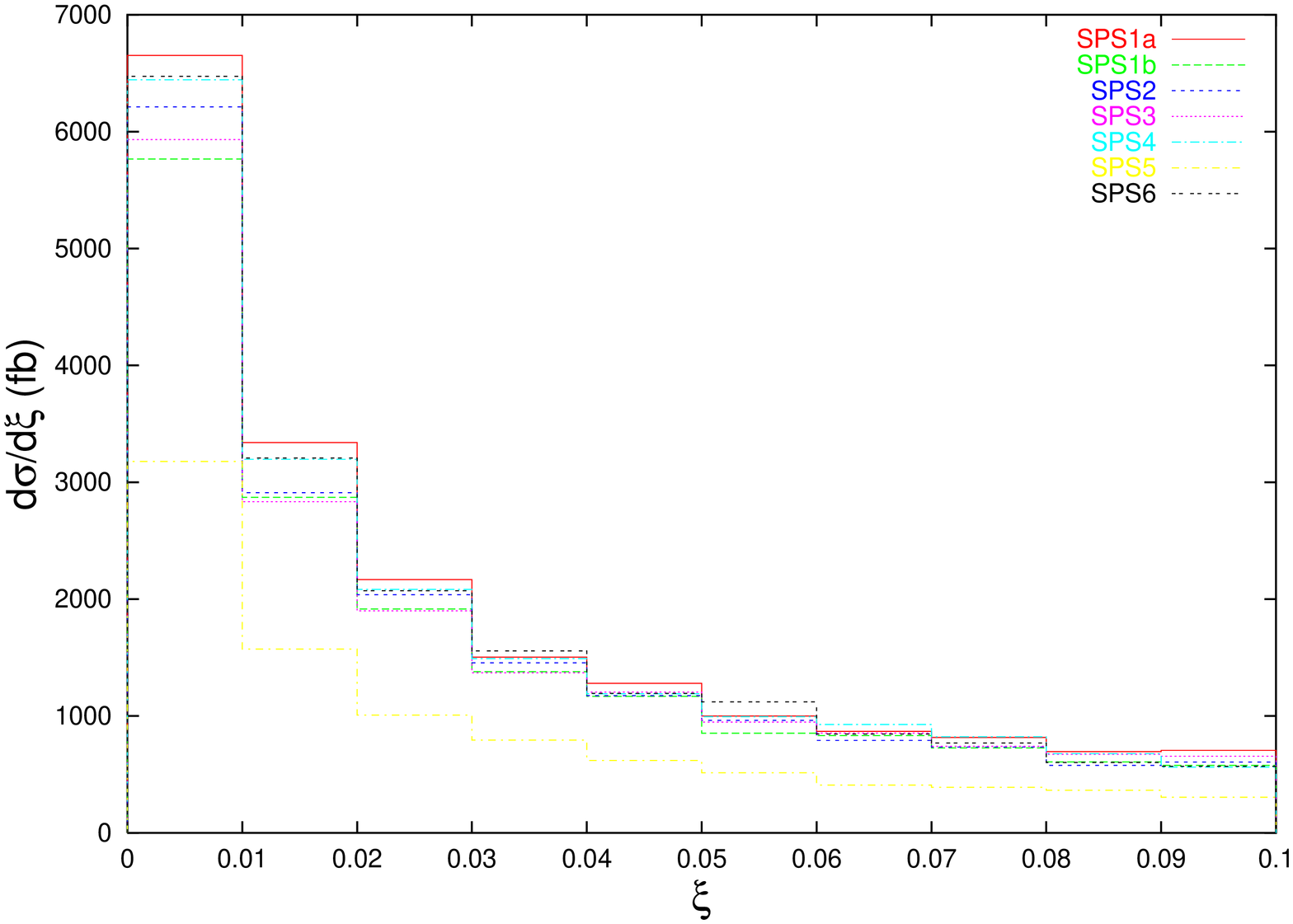}\includegraphics[width=4.8cm,
angle=-90]{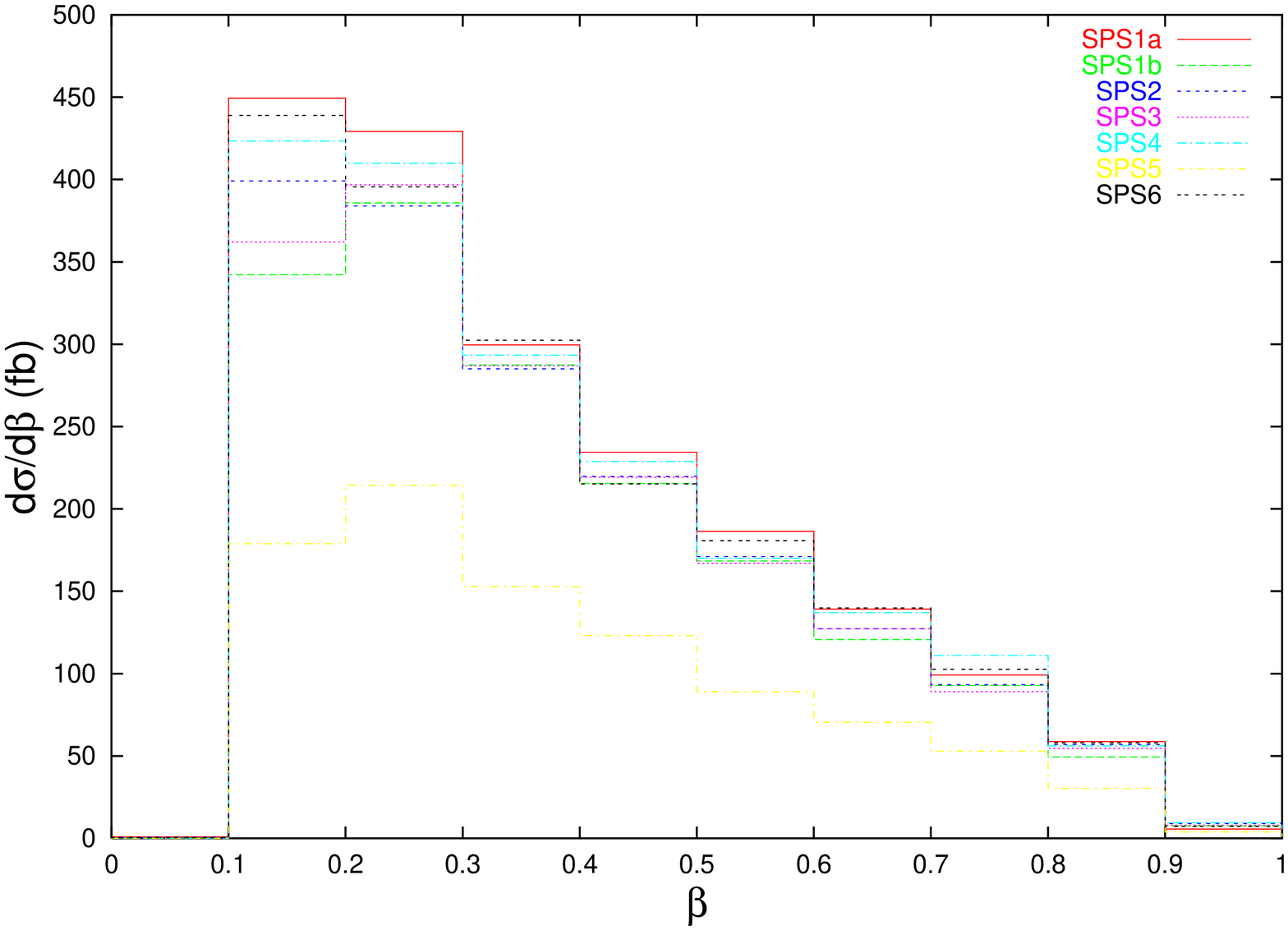} 
\caption{Differential cross--section for inclusive double diffractive
production of the lightest MSSM Higgs at the LHC for mSUGRA benchmark points.}
\end{center}
\end{figure}\\
\\
{\bf SUSY breaking in minimal supergravity (mSUGRA SPS 6)}\\
\\
This case corresponds to non-unified gaugino masses at the GUT scale 
with the bino having a 
mass parameter larger than previous mSUGRA models. The parameterisation is 
shown in Table~2. The differential and total inclusive double diffractive 
cross--section are not affected by these new values of the masses as can be 
observed in Table~2 and Fig.~2.
\begin{table}[h] 
\begin{center}
    \begin{tabular}{|c|c|c|c|c|c|c|c|c|c|}  \hline      
    Non--Universal SUGRA&$m_0$&$m_{1/2}$&$A_0$&$\tan{\beta}$&$m_1$&$m_{2,3}$&$m_{h^0}$&$\sigma_{~{\rm
P \hspace{-0.25cm}I}}$ (fb)\\
\hline \hline
    $6$&100&250&-100&10&480&300&115&184\\ 
    \hline   
    \end{tabular} 
\end{center}
\vspace{0.4cm}
\caption{Parameterisation of mSUGRA point SPS6 and total diffractive 
cross--section.}
\label{table2}
\end{table}\\
\\
{\bf Gauge-mediated SUSY breaking (GMSB SPS 7-8)}\\
\\
In this framework SUSY is also broken in a hidden sector and the mediation to 
the visible one is via gauge interactions. In the minimal case the parameters 
are now a universal 
soft SUSY breaking mass scale $\Lambda$, the messenger mass $m_{\rm mes}$ and 
index $N_{\rm mes}$, and the usual $\tan{\beta}$. The values for the parameters in the SPS 7 and 8 
points are indicated in Table~3, together with the large values for the 
cross--section. The differential cross--sections in $\xi$ and $\beta$ are 
shown in Fig.~3. 
\begin{table}[h] 
\begin{center}
    \begin{tabular}{|c|c|c|c|c|c|c|c|}  \hline  
    GMSB&$\Lambda$&$m_{\rm mes}$&$N_{\rm
mes}$&$\tan{\beta}$&$m_{h^0}$&$\sigma_{~{\rm P \hspace{-0.25cm}I}}$ (fb)\\
\hline \hline
    $7$&40000&80000&3&15&114&190\\ 
    $8$&100000&200000&1&15&115&182\\ 
\hline     
    \end{tabular}
\vspace{0.4cm}
\end{center} 
\caption{Parameterisation of GMSB points and total diffractive 
cross--sections.}
\end{table}\\
\\
{\bf Anomaly-mediated SUSY breaking (AMSB SPS 9)}\\
\\In this theoretical framework the SUSY breaking is mediated to the visible 
sector using the so--called super--Weyl anomaly. The values of the 
parameters are given in Table 4. Again, as in all the MSSM Snowmass points 
studied in this Letter, the inclusive 
cross--sections for the production of the lightest Higgs are large, see 
Table~4 and Fig.~4.
\begin{table}[h] 
\begin{center}
    \begin{tabular}{|c|c|c|c|c|c|}  \hline      
    AMSB&$m_0$&$m_{3/2}$&$\tan{\beta}$&$m_{h^0}$&$\sigma_{~{\rm P
\hspace{-0.25cm}I}}$ (fb)\\
\hline \hline
    9&450&60000&10&115&181\\ 
\hline      
    \end{tabular}
\end{center} 
\vspace{0.4cm}
\caption{Parameterisation of AMSB SPS 9 point and total diffractive 
cross--section.}
\end{table}
\begin{figure}
\begin{center} ~~~~~~GMSB\\
\includegraphics[width=4.8cm, angle=-90]{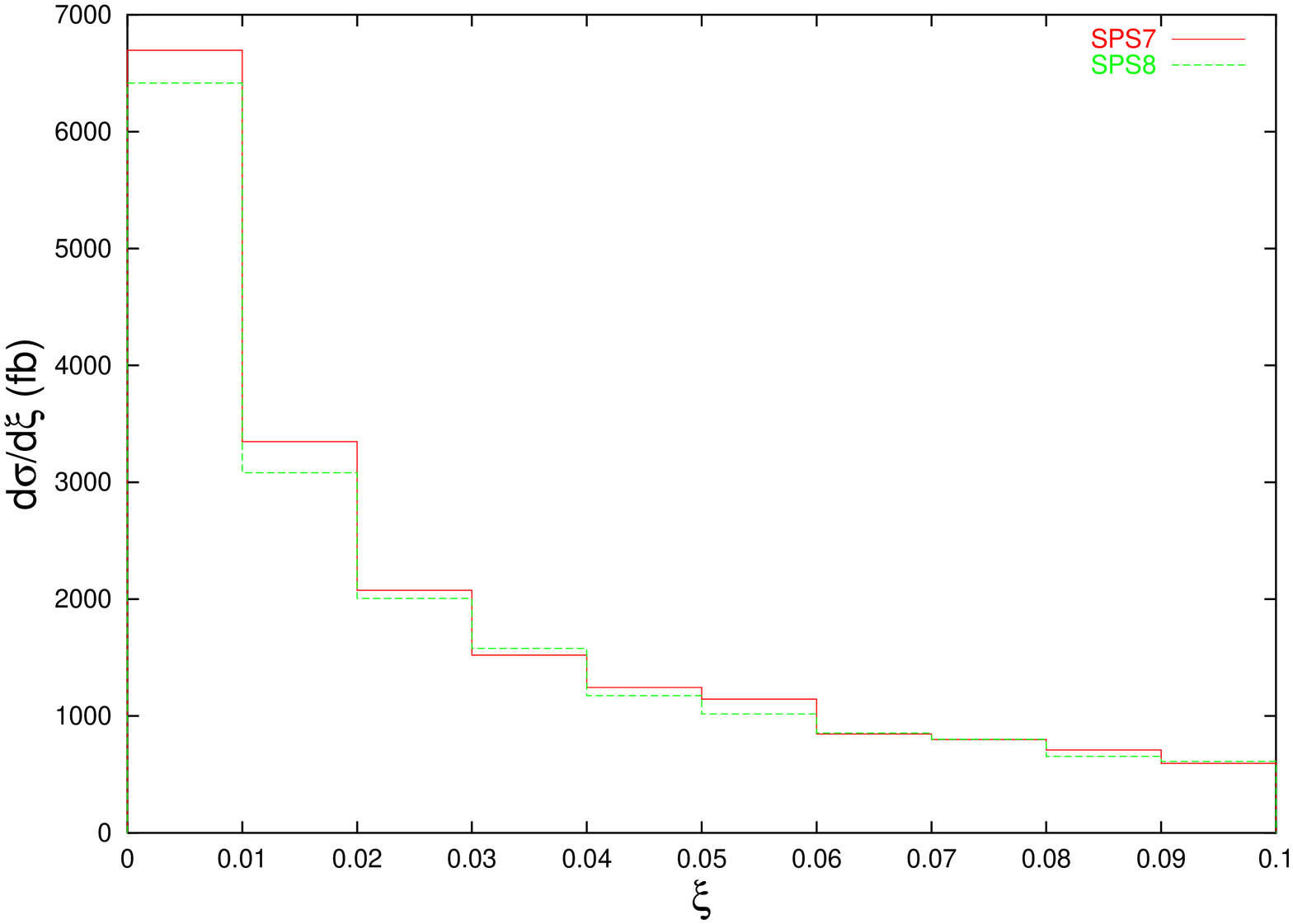}\includegraphics[width=4.8cm,
angle=-90]{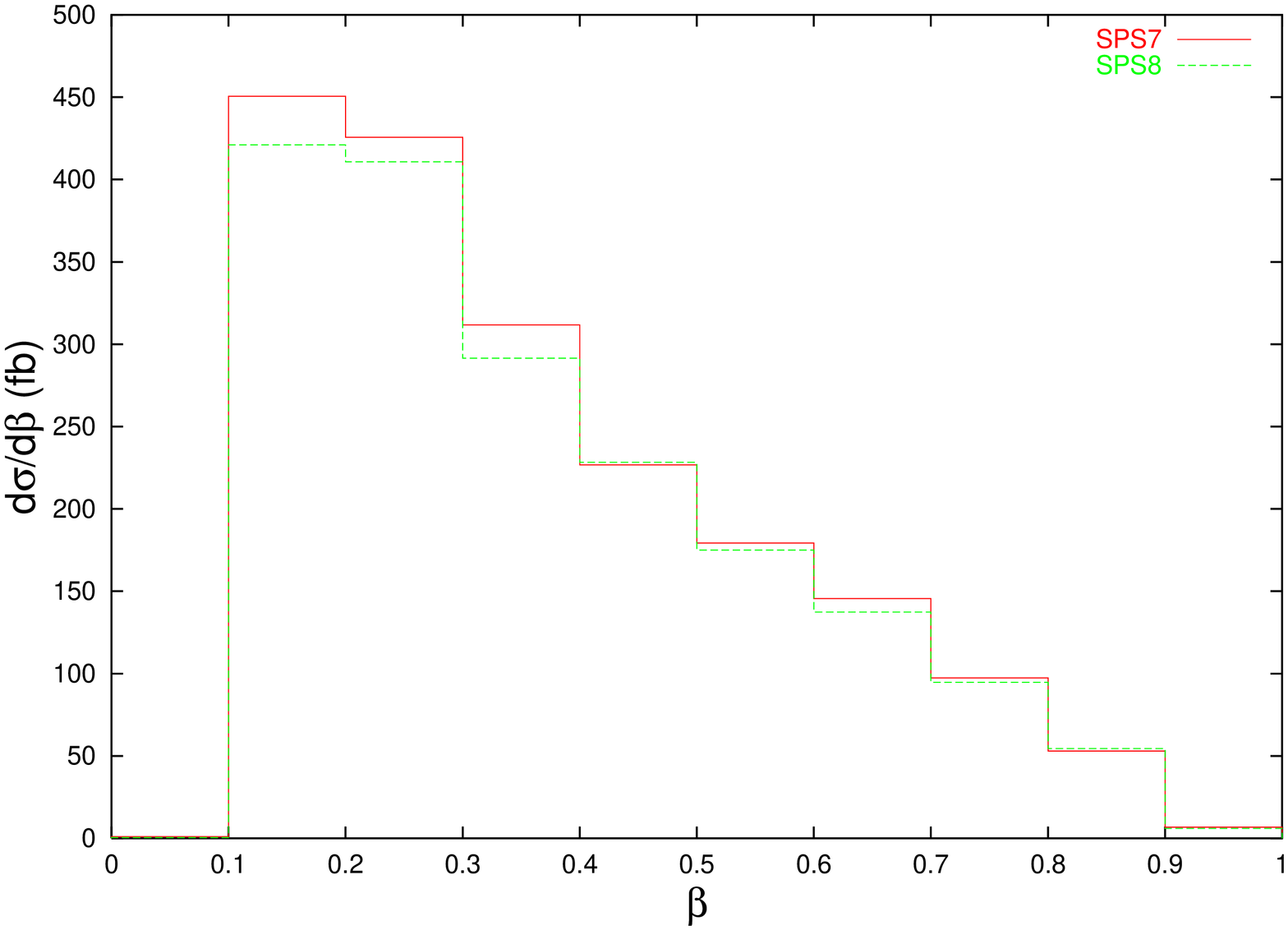}
\caption{Differential cross--section for inclusive double diffractive
production of the lightest MSSM Higgs at the LHC for GMSB benchmark points.}
\end{center}
\label{figSPS7}
\end{figure}
\begin{figure}
\begin{center} ~~~~~~AMSB\\
\includegraphics[width=4.8cm, angle=-90]{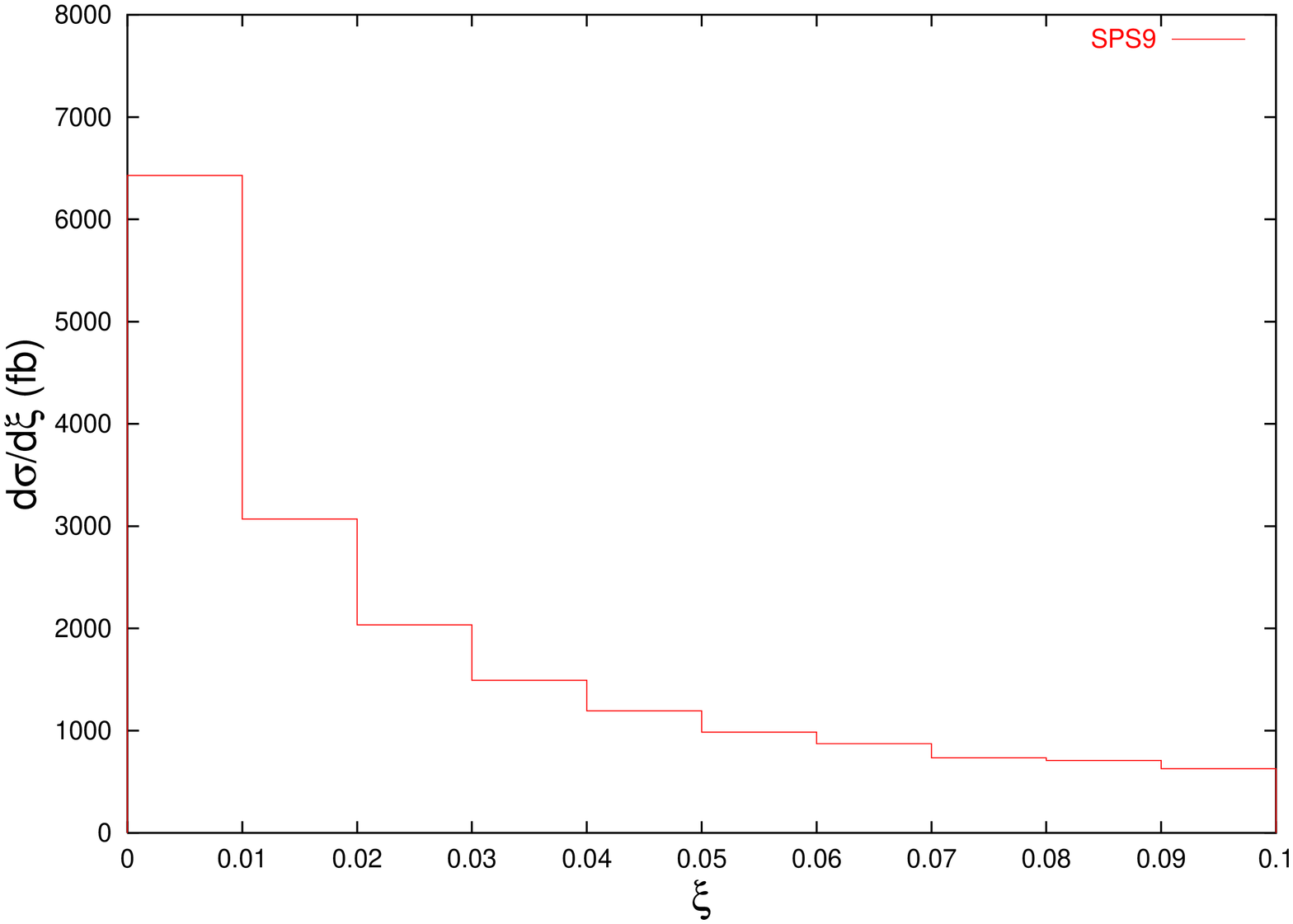}\includegraphics[width=4.8cm,
angle=-90]{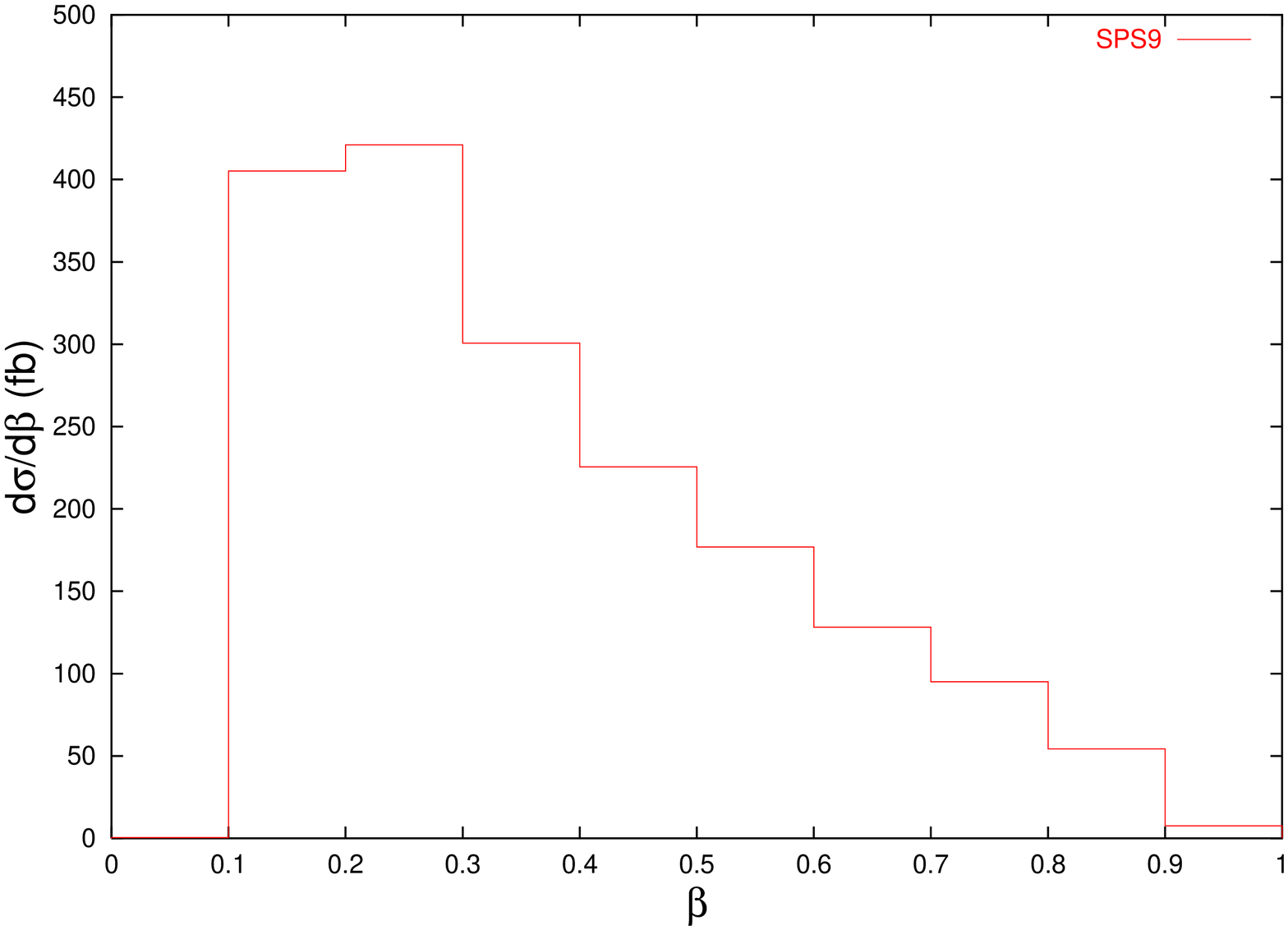}
\caption{Differential cross--section for inclusive double diffractive
production of the lightest MSSM Higgs at the LHC for AMSB benchmark point.}
\end{center}
\label{figSPS9}
\end{figure}

\section{Conclusions}

We have numerically estimated the inclusive double diffractive production 
of SUSY particles at the energies available at the future Large Hadron 
Collider at CERN. We have shown that the only cross--section large enough 
to provide a clean signal in this inclusive channel is that of the production 
of the lightest MSSM Higgs boson. This is the case provided the backgrounds are not too large, a point which will be investigated in a future work. 
Nevertheless, and always understanding that our results suffer from a 
theoretical uncertainty mainly due to the gap survival factor, the 
results are encouraging, showing large cross--sections 
for the inclusive double diffractive channel.  
It would also be interesting to investigate the production rates for SUSY 
particles in other channels, like single diffractive production, and exclusive 
processes, where there are no pomeron remnants in the final state.

\section*{Acknowledgements} 

We would like to thank Ben Allanach, Brian Cox, Jeff Forshaw and Bryan Webber 
for useful communications. A.S.V. acknowledges the 
support of PPARC (Postdoctoral Fellowship: PPA/P/S/1999/00446).

\end{document}